\documentclass[twocolumn,superscriptaddress,nopacs,aps,prl]{revtex4-2}
\usepackage{amssymb}
\usepackage{amsmath}
\usepackage{graphicx}
\usepackage{bm}
\usepackage{color}
\usepackage{xcolor}
\usepackage[normalem]{ulem}
\usepackage{physics}
\usepackage{subfigure}
\usepackage{natbib}
\usepackage{hyperref}

\graphicspath{%
    {converted_graphics/}
    {/}
}

\def\cF{\mathcal{F}_T}
\def\cM{\mathcal{M}}
\def\cR{\mathcal{R}}
\def\cT{\mathcal{T}}
\def\cS{\mathcal{S}}

\def\kf{f}
\def\mT{T}

\def\mB{\mathrm{B}}
\def\mD{\mathrm{D}}
\def\mmedium{\mathrm{m}}
\def\msphere{\mathrm{s}}
\def\meff{\mathrm{eff}}

\def\mfit{\mathrm{rm}}

\begin{document}

\title{Universal Casimir interaction between two dielectric spheres in salted water}

\author{Tanja Schoger}
\affiliation{Universität Augsburg, Institut für Physik, 86135 Augsburg, Germany}

\author{Benjamin Spreng}
\affiliation{Department of Electrical and Computer Engineering, University of California, Davis, CA 95616, USA}

\author{Gert-Ludwig Ingold}
\email{gert.ingold@physik.uni-augsburg.de}
\affiliation{Universität Augsburg, Institut für Physik, 86135 Augsburg, Germany}

\author{Paulo A. Maia Neto}
\email{pamn@if.ufrj.br}
\affiliation{Instituto de F\'{\i}sica, Universidade Federal do Rio de Janeiro \\ Caixa Postal 68528,   Rio de Janeiro,  RJ, 21941-972, Brazil}

\author{Serge Reynaud}
\email{serge.reynaud@lkb.upmc.fr}
\affiliation{Laboratoire Kastler Brossel, Sorbonne Universit\'e, CNRS, ENS-PSL, Coll\`ege de France, Campus Jussieu, 75005 Paris, France }

\date{\today}

\begin{abstract}
We study the Casimir interaction between two dielectric spheres immersed in a salted solution at distances larger than the Debye screening length. The long distance behavior is dominated by the non-screened interaction due to low-frequency transverse magnetic thermal fluctuations. It shows universality properties in its dependence on geometric dimensions and independence of dielectric functions of the particles, with these properties related to approximate conformal invariance.
\end{abstract}

\maketitle

The electromagnetic Casimir effect \cite{Casimir1948,Decca2011,Gong2021} and the so-called critical Casimir effect \cite{Fisher1978,Hertlein2008,Magazzu2019} are two examples of long-range forces appearing when fluctuations are confined within walls~\cite{Parsegian2006}. The former is often considered as associated to quantum field fluctuations and the latter to classical thermal fluctuations in matter. A problem sharing properties with both of them corresponds to the high-temperature limit of the electromagnetic Casimir interaction between bodies separated by a medium.

The case of two metallic spheres described by a Drude conductivity model in vacuum has been shown to lead to a universal expression in the limit of high temperatures or, equivalently, of large distances, with the free energy not depending on the details of the electromagnetic response of the involved material \cite{CanaguierDurand2012,Bimonte2012}.  
At room temperature $T\sim300\,$K, this universal thermal Casimir contribution overtakes the nonuniversal terms at large distances \cite{Bostrom2000} of the order of the thermal wavelength $\hbar c/k_\mB T\sim7.6\,\mu$m, making its experimental detection challenging as the magnitude of the force is decreasing as a power-law function of the distance \cite{Sushkov2011}.

We will consider in this letter a complementary case with two dielectric particles separated by a conducting electrolyte solution. In this case, there is also a universal classical expression in the limit of high temperatures, with the free energy not depending on the detailed dielectric function of the involved material. The universal thermal contribution now overtakes the sum of nonuniversal terms at a much smaller distance $\ell_\mT$ of the order of $0.1\,\mu$m for typical materials considered in this context \cite{Parsegian1971,MaiaNeto2019}. The existence and magnitude of this universal Casimir interaction has recently been confirmed by measurements \cite{Pires2021} involving a silica microsphere held by optical tweezers \cite{Ether2015} in the vicinity of a larger sphere, with both spheres immersed in salted water.

The high-temperature Casimir interaction between dielectric spheres in a conducting solution is a universal function of the geometry that can be derived exactly with the help of analytical and numerical tools based on the scattering approach \cite{Lambrecht2006} by using the plane-wave basis \cite{Spreng2020}. We present below the universal interaction free energy obtained in this manner and show that it can reach the thermal energy scale $k_\mB T$ and thus have important potential applications for the physics of biological interfaces and colloidal solutions \cite{Israelachvili2011}. 


In the following, we study the Casimir interaction between two dielectric spheres immersed in a salted solution at ambient temperature under the assumptions that the distance of closest approach $L$ between the spheres is much larger than the Debye screening length $\lambda_\mD$ and larger than the length $\ell_\mT$ introduced above. The first assumption implies that electrostatic interactions resulting from surface charges or electric potential fluctuations \cite{Mitchell1974,MahantyNinham1976,Nunes2021} are efficiently screened while the second assumption ensures that the force is dominated by the Matsubara term at zero frequency \cite{Schwinger1978}. Under these assumptions, the main force is due to the Casimir interaction mediated by low-frequency transverse magnetic thermal fluctuations coupled to electric multipoles. 

We have calculated numerically the exact free energy for the geometry of two spheres with arbitrary radii by adapting the plane-wave approach introduced in \cite{Spreng2020} to our problem. 
It turns out that the free energy shows universality properties for arbitrary values of geometric parameters and, in particular, that the interaction does not depend on the dielectric functions of the spheres. 
We first recall the method for calculating the free energy and then give approximated expressions allowing one to obtain simple estimates for it without having to perform the complete numerical calculation. 

The free energy $\cF$ in the high-temperature limit can be written as the product of an energy scale, the thermal energy $k_\mB T$, and a dimensionless function $\kf_u$ depending only on two ratios of the geometric dimensions $L,R_1,R_2$
\begin{equation}
\cF(L,R_1,R_2)=-k_\mB T \,\kf_u  ~.
\label{eq:defu}
\end{equation}
The negative sign of $\cF$ implies attraction between the spheres. As $\kf_u$ does not depend on $T$, the entropy is $\cS=k_\mB \,\kf_u$ and the free energy is $\cF=-T \cS$, so that the Casimir force is purely entropic in nature here. 

The parameter used as a subscript in $f_u$ measures the ratio of radii
in an expression symmetric in $R_1,R_2$
\begin{equation}
u=\frac{R_1R_2}{\left(R_1+R_2\right)^2}  ~.
\end{equation}
It lies in the range $0\leq u\leq\frac14$, with $u=0$ corresponding to a plane-sphere geometry and $u=1/4$ to two spheres of equal radii. A natural choice for the other parameter could be the dimensionless distance 
\begin{equation}
 x=\frac{L}{R_\meff}~,
\end{equation}
where $R_\meff=R_1R_2/(R_1+R_2)$ is the effective radius of the system of two spheres.
We will see however that a better parameter for describing the dependence of $\kf_u$ on distance is 
\begin{equation}
y=\frac{(L+R_1+R_2)^2-R_1^2-R_2^2}{2R_1R_2}=1+x+u\frac{x^2}{2}  ~. 
\label{eq:defy}
\end{equation}
This geometric quantity $y$ is invariant under conformal transformations generated by isometries and inversion in 3-dimensional Euclidean space \cite{Bromwich1900}. It has been known for a long time \cite{Thomson1853,Maxwell1873} to simplify the expression of the mutual capacitance $C_{12}$ between the two spheres ($C_{12}$ is written in terms of $\varpi=\text{arcosh}(y)$ in \cite{Maxwell1873}, Part 1 \S 11).

\begin{figure}[t!]
 \includegraphics[width=0.86\columnwidth]{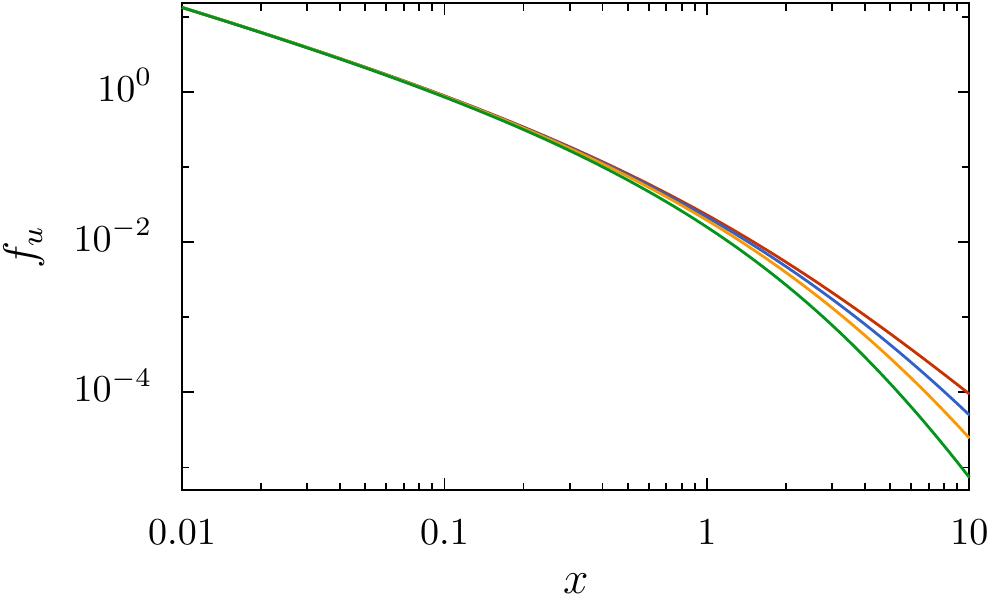}
  \caption{Reduced free energy $\kf_u$ for two dielectric spheres in salted water, drawn as a function of $x=L/R_\meff$ for different values of $u=0,0.04,0.10,0.25$ from top to bottom. }
\label{fig:plotfuvsx}
\end{figure}

Within the scattering approach, $\kf_u$ can be written in terms of an operator $\cM$ representing the effect on the electromagnetic field of a single round-trip in the cavity formed by the two spheres 
\begin{equation}
\kf_u =-\frac{\Tr\log\left(1-\cM\right)}2  
= \sum_{r=1}^\infty \kf_u^{(r)} ~,\quad
\kf_u^{(r)} \equiv \frac{\Tr \cM^r}{2r}  ~.
\label{eq:sumoverroundtrips}
\end{equation}
The contribution $\kf_u^{(r)}$ corresponds to a given number $r$ of round-trips in the cavity, and $\kf_u$ is the sum over all numbers of round-trips. The round-trip operator $\cM$ is defined as a product of reflection operators ${\cal R}_m$ for the spheres $m=1,2$ and translation operators ${\cal T}_{m'm}$ from a frame aligned on sphere $m$ to the one aligned on $m'$
\begin{equation}
\cM= \cR_1\,\cT_{12}\, \cR_2\,\cT_{21}~.
\end{equation}
These operators can be expressed explicitly in the basis of plane waves \cite{NietoVesperinas2006} characterized by the projection $\mathbf{k}$ of the wave vector onto the plane perpendicular to the line joining the two centers of the spheres with the direction of propagation changing at each reflection.

In this basis, the translation operators are diagonal 
\begin{equation}
 \langle\mathbf{k}'\vert\cT_{m'm}\vert\mathbf{k}\rangle = e^{-k(L+R_1+R_2)}\delta^{(2)}(\mathbf{k}'-\mathbf{k})\,,
\end{equation}
where $k=\vert\mathbf{k}\vert$ is the norm of the projected wave vector $\mathbf{k}$. The reflection operators $\cR_1$ and $\cR_2$ can be derived by following \cite{Spreng2018} and in particular
its Appendix~B where the zero-frequency limit is described. When considering spheres in salted water instead of vacuum as was done in \cite{Spreng2018}, we need to replace the static dielectric constant $\varepsilon(0)$ by the ratio $\varepsilon_\msphere(0)/\varepsilon_\mmedium(0)$ of static dielectric functions of spheres and medium.

\begin{figure}[t!]
 \includegraphics[width=0.86\columnwidth]{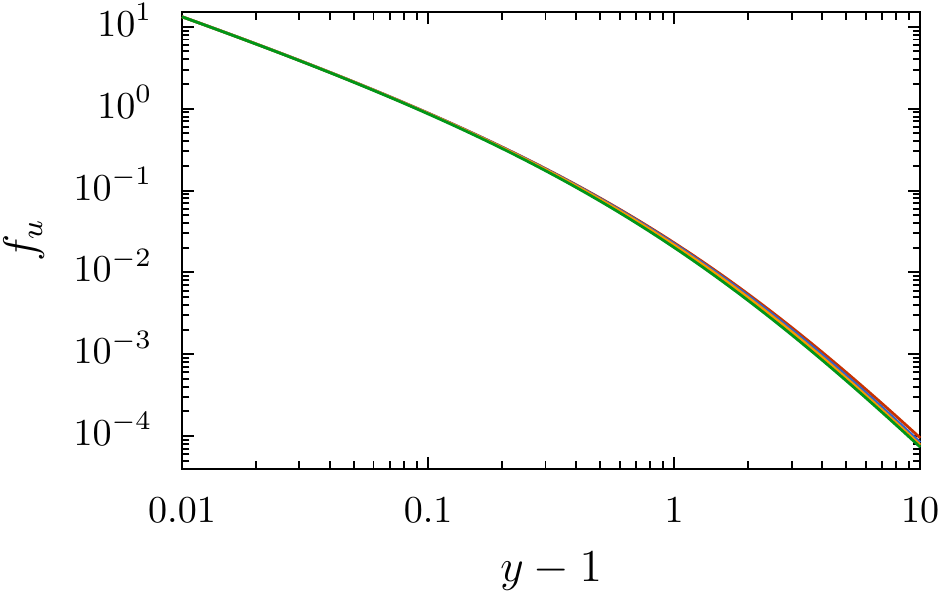}
 \caption{Reduced free energy $\kf_u$ with the same conventions as on Fig.~\ref{fig:plotfuvsx}, but for the use of abscissa $y-1$. The range of abscissas matches that on Fig.~\ref{fig:plotfuvsx} for the curve $u=0$.}
\label{fig:plotfuvsym}
\end{figure}

Using salted water as the medium has two consequences for our problem. Moving ions efficiently screen longitudinal modes \cite{MaiaNeto2019} and we can restrict our attention to transverse magnetic modes. Furthermore, in view of the finite static conductivity of the medium, $\varepsilon_\mmedium(0)$ is infinite, so that any detailed dielectric property of the particles disappears from the matrix elements of the reflection operators $\cR_1$ and $\cR_2$
\begin{equation}
 \langle\mathbf{k}'\vert\mathcal{R}_m\vert\mathbf{k}\rangle =
 -\frac{2\pi R_m}{k'}\sum_{\ell=1}^\infty\frac{\ell}{\ell+1}\frac{\left(2R_m^2kk'(1+\cos\varphi)\right)^\ell}{(2\ell)!}
 \,, \label{reflectionoperators}
\end{equation}
with $\varphi$ the angle between the in- and out-going projected wave vectors $\mathbf{k}$ and $\mathbf{k}'$.

Our numerical results for $\kf_u$ are shown in  Figs.~\ref{fig:plotfuvsx} and \ref{fig:plotfuvsym} as functions of $x$ and $y-1$, respectively, for fixed values of $u$, \emph{i.e.}, fixed ratios of radii.
Both plots show a monotonic decrease of free energy from small to large distances $L$.
Their comparison reveals that the dependence of $\kf_u$ on $u$ seen on Fig.~\ref{fig:plotfuvsx} is largely captured on Fig.~\ref{fig:plotfuvsym} by using the abscissa $y-1=x(1+\frac{ux}{2})$ which is a stretched version of $x$ with the stretching factor depending on $u$. 
The fact that the different curves $\kf_u$ are better aligned on Fig.~\ref{fig:plotfuvsym} will be given an interpretation by the analytical results presented below. 

\begin{figure}[t!]
\includegraphics[width=0.86\columnwidth]{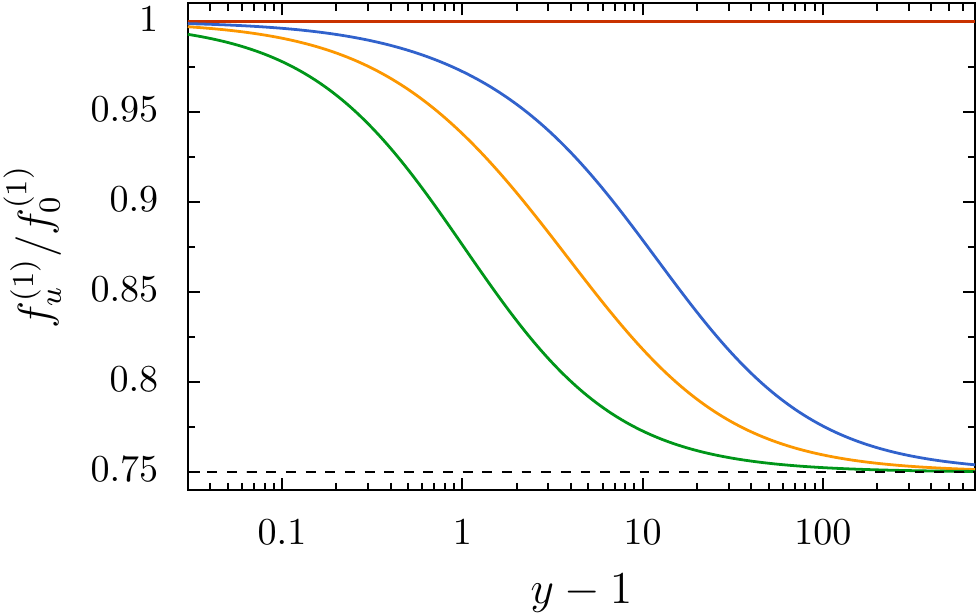}
\caption{Ratio $\kf_u^{(1)}/\kf_0^{(1)}$ versus $y-1$ for different values of $u=0,0.04,0.10,0.25$ from top to bottom. All curves for $u\neq0$ go from the value 1 at $y-1\ll1$ (top line, $u=0$) to the value $\tfrac34$ at $y-1\gg1$ (bottom dashed line). }
\label{fig:ratiofonf0}
\end{figure}

Making use of the techniques employed in \cite{Schoger2021}, the analytical expression for the single round-trip contribution can be written as
\begin{eqnarray}
\kf_u^{(1)} &=& \frac{y}{4 \left(y^2-1\right)}
+\frac{z}{12}  \log \frac{\left(y^2-1\right) z^2}{\left(y z+\frac{1}{2}\right)^2}     \label{eq:analyticalf1}   \\
&+&\frac{1}{12 \sqrt{z}}\sum_{\eta=\pm}
\frac{1}{\alpha_\eta ^{3/2}} 
\log \frac{2 y^2+\alpha_\eta y-1+\sqrt{\alpha_\eta  z}}{2 y^2+\alpha_\eta y-1-\sqrt{\alpha_\eta z}}   ~, \notag 
\end{eqnarray}
where auxiliary variables have been introduced 
\begin{equation}
\alpha_\pm=\frac{1-2u\pm\sqrt{1-4 u}}{2 u}
~,\quad z=2y+\sum_{\eta=\pm}\alpha_\pm ~. \label{eq:defalpha}
\end{equation}
The function $\kf_u^{(1)}$ mainly depends on $y$ which means that the analytical results nearly obey conformal invariance. But $\kf_u^{(1)}$ still depends on $u$ through the parameters $\alpha_\pm$ which correspond to the ratios $R_1/R_2$ and  $R_2/R_1$, and this dependence breaks exact invariance. 

The large-distance limit, $L\gg R_\meff$ is dominated by the single round-trip contribution obtained from the asymptotic expansion of (\ref{eq:analyticalf1}) 
\begin{eqnarray}
 \kf_{u=0} \simeq \frac{1}{8y^3} ~,\quad 
 \kf_{u\neq0} \simeq \frac{3}{32y^3} 
 \simeq \frac{3}{4} \kf_{u=0} ~.
 \label{eq:limdip}
\end{eqnarray}
The free energy in the sphere-sphere geometry ($u\neq0$) is smaller by a factor $\tfrac34$ relative to the plane-sphere geometry ($u=0$). This factor is the main reason for the dependence of $\kf_u$ on $u$ observed at large values of $y-1$ in Fig.~\ref{fig:plotfuvsym}, that is the breaking of exact conformal invariance.

While the asymptotic power-law dependence on $y$ in (\ref{eq:limdip}) is the same for all values of $u$, this is no longer the case when the free energy is expressed as a function of $x$. Then, the free energy in the plane-sphere case decreases as $x^{-3}$ while it decays as $(ux^2)^{-3}$ for two spheres, thus explaining why the asymptotics is so different in Fig.~\ref{fig:plotfuvsx}. In simple words, the asymptotic behaviors \eqref{eq:limdip} explain why using the abscissa $y-1$ captures most of the distance dependence of the free energy while still weakly breaking exact conformal invariance. 

In the short-distance limit $L\ll R_\meff$, multiple round-trips need to be accounted for. In this limit, $\kf^{(1)}_u$ is independent of $u$ and the same holds for multiple round-trips with $\kf_u^{(r)} \simeq \kf_u^{(1)}/r^3$. It follows that the sum over $r$ can be written as a simple factor, the Apéry's constant $\zeta(3)$,
\begin{eqnarray}
 \kf_u \simeq \sum_{r=1}^\infty\frac{\kf_u^{(1)}}{r^3} \simeq \frac{\zeta(3)}{8\left(y-1\right)}  ~.
 \label{eq:limpfa}
\end{eqnarray}
This result explains why the curves $\kf_u$ calculated for different values of $u$ tend to become identical when $x\ll1$ in Fig.~\ref{fig:plotfuvsx} and $y-1\ll1$ in Fig.~\ref{fig:plotfuvsym}. 
The free energy (\ref{eq:limpfa}) also corresponds to the so-called proximity-force approximation, where the force can be obtained by integrating the pressure calculated between two planes over the range of distances met in the geometry of two spheres. 

\begin{figure}[t!]
\includegraphics[width=0.86\columnwidth]{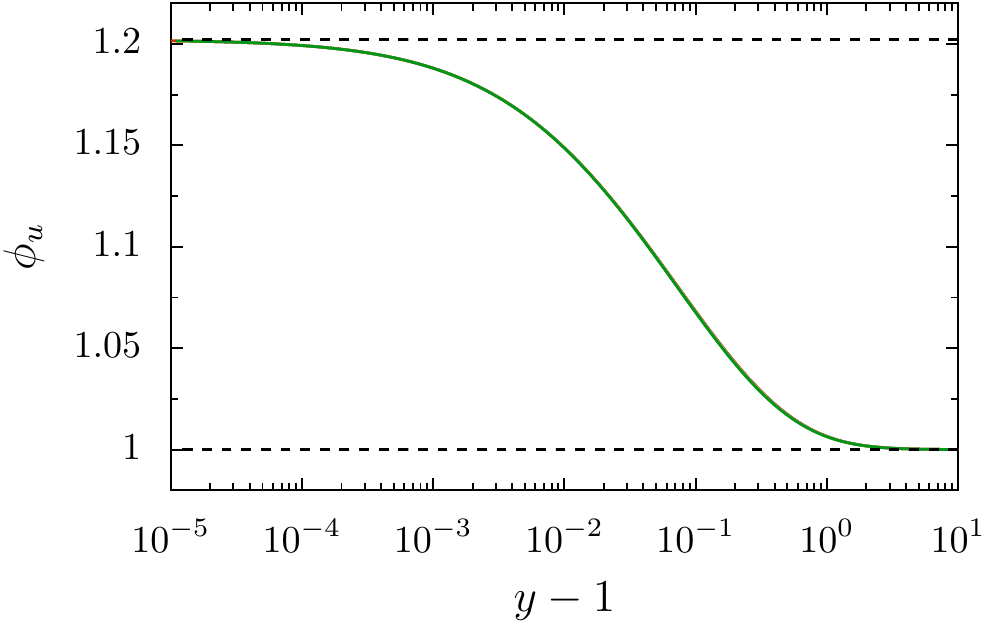}
\caption{Ratio $\phi_u \equiv \kf_u/\kf_u^{(1)}$ with the same conventions as on Fig.~\ref{fig:plotfuvsym}, except that a linear scale is used for the ordinate. The curves for different values of $u$ are practically indistinguishable from each other. They produce a universal function decreasing from $\zeta(3)\simeq1.202$ at small $y-1$ (top dashed line) to 1 at large  $y-1$  (bottom dashed line). }
\label{fig:phiuvsYm}
\end{figure}

A remarkable fact appears when drawing the ratio of the full expression $\kf_u$ to the single round-trip one $\kf_u^{(1)}$
\begin{eqnarray}
 \phi_u (y) \equiv \frac{\kf_u(y)}{\kf_u^{(1)}(y)}   ~.
 \label{eq:defphi}
\end{eqnarray}
As expected from previous discussions, this ratio goes from the constant $\zeta(3)\simeq1.202$ at small values of $y-1$ to the constant 1  at large values of $y-1$. We see on Fig.~\ref{fig:phiuvsYm} the even stronger property that the ratio $\phi_u$ is a monotonically decreasing function of $y-1$ depending very weakly on the parameter $u$. 
Precisely, the curves drawn on Fig.~\ref{fig:phiuvsYm} with the same conventions as on Fig.~\ref{fig:plotfuvsym} are practically indistinguishable from each other, which corresponds to a nearly exact conformal invariance. 
The very weak dependence on $u$ is assessed by calculating numerically the ratio $\phi_u/\phi_\star$ with $\phi_\star$ evaluated for a fixed value $u=u_\star$. Choosing for $u_\star$ the value $u=0.1$ minimizes the deviation of $\vert\phi_u/\phi_\star-1\vert$ which remains smaller than $4\times10^{-4}$ on the whole domain of parameters $y>1$ and $0\leq u\leq\tfrac14$. 

The reason for this nice universality property can be understood qualitatively. The contributions of multiple round-trips are important when the single round-trip ones are themselves large, that is in the domain $y-1\ll1$ where all contributions tend to become independent of $u$.
A significant dependence of contributions on $u$ would appear in the opposite domain but $\phi_u$ is anyway close to unity there since $\kf_u$ and $\kf_u^{(1)}$ tend to become identical. These simple arguments explain why the dependence of $\phi_u$ on $u$ remains weak.

It is furthermore possible to find a rational function of the argument $e^{y-1}$ which fits the different curves $\phi_u$ with a small maximal error on the whole domain of parameters
\begin{equation}
\left\vert\frac{\phi_u(y)}{\phi_\mfit(y)}-1\right\vert<\epsilon ~.
\end{equation}
The rational model function $\phi_\mfit$ has the following form 
\begin{equation}
 \phi_\mfit(y) =\prod_{k=1}^n\frac{e^{y-1}-1+\nu_k}{e^{y-1}-1+\mu_k } ~,
 \label{eq:rationalmodel}
\end{equation}
where $\nu_k,\mu_k$ are roots of the polynomials in the numerator and denominator. They have been written as values close to 0, with all values $\nu_k,\mu_k$ being positive to avoid large deviations on the domain $y>1$.
Inserting the coefficients given in Table \ref{table:coefficients} in the rational function \eqref{eq:rationalmodel} leads to a maximal deviation $\epsilon\simeq1.2\times10^{-3}$. 

\begin{table}[b!]
\begin{tabular}{|c|c|c|}
\hline
~ & $k=1$ & $k=n=2$ \\
\hline
~ $\nu_k$ ~ & ~ 0.004618 ~ & ~ 0.09639 ~ \\
\hline
~ $\mu_k$ ~ & ~ 0.004415 ~ & ~ 0.08397 ~ \\
\hline
\end{tabular}
\caption{Table of coefficients to be used in the model \eqref{eq:rationalmodel} for $n=2$ with maximal deviation $\epsilon=1.2\times10^{-3}$. 
}
\label{table:coefficients}
\end{table}

We are left in the end of this reasoning with a largely simplified expression of the full function $\kf_u(y)$ 
\begin{equation}
 \kf_u(y) = \kf_u^{(1)}(y) \, \phi_\mfit(y)  ~.
 \label{eq:approx}
\end{equation}
The first factor is the analytical single round-trip expression \eqref{eq:analyticalf1} depending on $y$ and $u$ while the second factor $\phi_\mfit$ is the rational function \eqref{eq:rationalmodel} with the associated parameters given in table \ref{table:coefficients}. 
This leads to an accuracy which should be sufficient for most applications
\footnote{Should a better accuracy be needed, a lower $\epsilon$ could be obtained with an higher order $n$ in the model \eqref{eq:rationalmodel}. For example, $n=4$ leads to a maximal error matching the level of deviations between different curves $\phi_u$. }.

As already mentioned, analogous universality properties were discussed for the Casimir free energy between metallic spheres in vacuum described by the Drude model at the high-temperature limit 
\cite{CanaguierDurand2012,Bimonte2012}. There, the reason for the universality was that the dielectric function of the spheres tends to infinity at zero frequency, due to the finite static conductivity of metals. In this sense, the problem studied here is dual of the preceding one, with the medium, salted water, exhibiting a finite static conductivity. This changes the numerical sum to be evaluated (the factor $\ell/(\ell+1)$ in Eq.~\eqref{reflectionoperators} has to be replaced by unity in the problem with metallic spheres \cite{Schoger2021}) and then the detailed results while preserving most qualitative discussions. A significant difference comes from the evaluation of the minimal distance for which the free energy is mainly given by the zero-frequency Matsubara term. In the case of vacuum, this distance is the thermal wavelength $7.6\,\mu$m whereas a much smaller distance $\ell_\mT\sim 0.1\,\mu$m is found in the case of salted water \footnote{The non-zero Matsubara terms can be evaluated by using the results in \cite{Spreng2020}; $\ell_\mT$ depends on the ratios of dielectric functions of spheres and water at non-zero Matsubara frequencies; the number given in the text is a maximum value for cases studied in \cite{Parsegian1971,MaiaNeto2019,Pires2021}.}. The universal expression studied in the present letter describes the force accurately on a much broader distance range than the analogous expression for metallic spheres in vacuum. Furthermore, there exists a wide range of distances where the Casimir force is sufficiently strong to make the universal thermal contribution of the electromagnetic field experimentally accessible \cite{Pires2021}.

In the end, the results of this letter imply that two spherical objects approaching each other in salted water with strong screening undergo a non-screened Casimir interaction having a universal dependence on geometric parameters, but no dependence on dielectric properties of the spheres. This interaction should have important potential consequences for the physics of biological interfaces and colloids as soon as it represents a significant fraction of the thermal energy $k_\mB T$ as the immersion in water imposes a Brownian motion to the spheres. This condition is met when the distance $L$ is of the order or smaller than one tenth of the effective radius (see Figs.~\ref{fig:plotfuvsx} and \ref{fig:plotfuvsym}). 
In the associated domain $y-1\lesssim 0.1$, the interaction is mainly determined by the conformally invariant parameter $y$ characterizing the two-spheres geometry, so that the breaking of exact conformal invariance through a dependence on $u$ remains weak (see Fig.~\ref{fig:ratiofonf0}). The interaction cannot be deduced precisely from any of the two limiting cases \eqref{eq:limdip} or \eqref{eq:limpfa}, but its estimation is easily deduced from Eq.~\eqref{eq:approx} with the function $\phi_\mfit$ having significant variations in the domain of interest for biological interfaces (see Fig.~\ref{fig:phiuvsYm}).

\begin{acknowledgments}
P.A.M.N. thanks Sorbonne Universit\'e for hospitality and
acknowledges funding from the Brazilian agencies 
Conselho Nacional de Desenvolvimento Cient\'{\i}fico e Tecnol\'ogico (CNPq--Brazil),
Coordena\c{c}\~ao de Aperfei\c{c}oamento de Pessoal de N\'{\i}vel Superior (CAPES--Brazil),  
Instituto Nacional de Ci\^encia e Tecnologia Fluidos Complexos  (INCT-FCx), and the
Research Foundations of the States of Rio de Janeiro (FAPERJ) and S\~ao Paulo (FAPESP). 
\end{acknowledgments}

\end{document}